\documentclass[preprint,aps]{revtex4}

\usepackage{graphicx}
\usepackage{dcolumn}
\usepackage{bm}

\begin{document}
\title{Surface-mode microcavity}
\author{Sanshui Xiao}
\altaffiliation[Also at ]{the Center for Optical and
Electromagnetic Research, Zhejiang University, YuQuan, Hangzhou,
310027, P. R. China.}
\author{Min Qiu}
\email{min@imit.kth.se} \affiliation{Laboratory of Optics,
Photonics and Quantum Electronics, Department of Microelectronics
and Information Technology, Royal Institute of Technology (KTH),
Electrum 229, 16440 Kista, Sweden}

\date{\today}

\begin{abstract}
Optical microcavities based on zero-group-velocity surface modes
in photonic crystal slabs are studied. It is shown that high
quality factors can be easily obtained for such microcavities in
photonic crystal slabs. With increasing of the cavity length, the
quality factor is gradually enhanced and the resonant frequency
converges to that of the zero-group-velocity surface mode in the
photonic crystal. The number of the resonant modes with high
quality factors is mainly determined by the number of surface
modes with zero-group velocity.

\end{abstract}

 \maketitle



Optical microcavities have attracted much attention due to their
potential applications in many various fields, such as ultrasmall
optical filters, high efficiency light emission diodes, low
threshold lasers, nonlinear optics and quantum information
precessing
\cite{Noda2000P1,Song2005P1,Qiu2003P2,Paiter1999P1,Spillane2002P1,Michler2000P1}.
All applications require microcavities to confine light strongly
and densely, that is, microcavities should have both a
high-quality factor $Q$ and small modal volume $V$. Total internal
reflection and/or Bragg reflection are generally used for light
confinement. For the smaller cavity, $Q$ is greatly reduced. One
of the best approaches to resolve the problem is the extension of
the Bragg reflection in a two-dimensional (2D) or a
three-dimensional (3D) photonic crystal (PhC) \cite{Noda2000P2}.
It has been also shown that extremely small-mode-volume high-$Q$
microcavities can be realized by 2D PhC slabs
\cite{Song2005P1,Yoshie2001P1,Akahane2003P1,Ryu2003P1,Akahane2003P2,Qiu2004P2},
owing to strong optical confinement for both the in-plane and
vertical direction.

Surface waves are propagating electromagnetic waves, which are
bound to the interface between materials and free space. For
dielectric interfaces, they usually do not exist on dielectric
materials. However, due to the existence of photonic band gaps, it
has been shown that dielectric PhCs may support surface waves for
some cases, e.g., a truncated or deformed structure at the
interface \cite{Joannopoulos}. In the present letter, we study PhC
microcavities utilizing surface modes with zero-group velocity.
Though the concepts apply for all PhCs, for simplicity, only
surface resonant modes in 2D PhC slabs are considered in this
letter. These cavities are open cavities in the sense that one of
the in-plane boundaries is exposed to exterior. Although the
similar structure has been realized by Yang et. al. experimentally
\cite{Yang2004}, the dependence of the quality factor and resonant
frequency to the cavity length has not been analyzed yet.

\begin{figure}[h]
\includegraphics[width=8 cm]{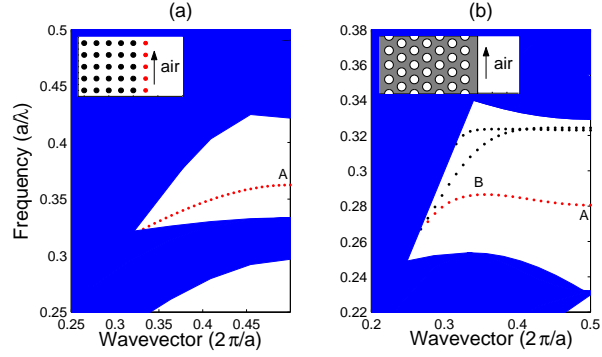}
\caption{\label{Fig1} (Color online) (a) The surface band
structure for the (10) surface of a square PhC slab of dielectric
rods in air, the radius of rods is $0.20a$ ($a$ is the lattice
constant), the height of the rods is $2.0a$, and the dielectric
constant of the rods is $11.56$. The radius of the rods at the
surface is $0.15a$. (b) The surface band structure for the $\Gamma
K$ surface of a triangular PhC slab of air holes extending through
a high-index ($\varepsilon=11.56$) finite-height dielectric slab,
the radius of holes is $0.3a$, and the high-index slab is of
thickness $0.6a$. The distance between the centers of first right
holes and the right boundary of the dielectric is $d=\sqrt{3}a
/2$. }
\end{figure}
\begin{figure}[h]
\includegraphics[width=6 cm]{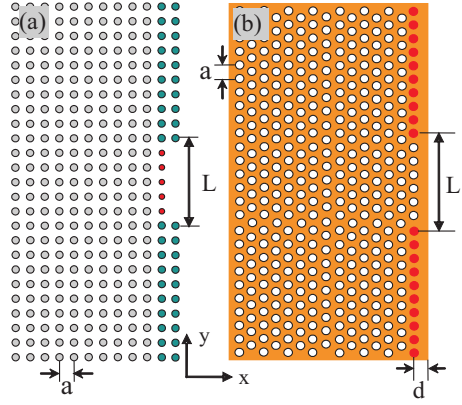}
\caption{\label{Fig2} (Color online) (a) Top view of the
microcavity composed by a 2D square PhC slab. (b) Top view of the
microcavity composed by a 2D triangular PhC slab. The length of
the cavities is denoted by $L$.}
\end{figure}

Let us first consider a 2D square PhC slab with dielectric rods in
air. The permittivity of the rods is $\varepsilon=11.56$, the
height of the rods is $2a$, and the radius of the rods is
$R=0.2a$, where $a$ is the lattice constant. Surface defects at
the interface are introduced by reducing the radius of rods
($R_d=0.15a$), the top view of which is illustrated in the inset
of Fig. \ref{Fig1} (a). The dispersion relation for the
reduced-radii rod-slab transverse magnetic (TM) surface modes is
calculated using the 3D finite-difference time-domain (FDTD)
method \cite{TafloveFDTD} and is shown in Fig. \ref{Fig1} (a),
where the shadow regions are the projected band structure for TM
modes. It can be seen from Fig. \ref{Fig1} (a) that such a surface
structure only supports one surface mode. Group velocity governed
by $v_g=\nabla_k \omega$ for the surface mode is zero for the wave
vector $k=\pi/a$ (Mode A). Suppose an optical microcavity that is
composed by the 2D square PhC slab, as shown in Fig. \ref{Fig2}
(a). The grey rods, acting as reflecting mirrors in the $y$
direction, are the same as those interior rods. The length of the
cavity is denoted by $L$. Surface modes can go through the central
surface but are terminated by the reflecting mirror. Only the mode
for obeying the Fabry-Perot condition (the round-trip accumulated
phase $\Phi=2kL+2\Delta\phi$ is a multiple of $2\pi$, where
$\Delta\phi$ is the phase shift associated with reflection from
the boundary) will stay in the cavity for a long time, i.e.,
become a surface resonant mode. For such a microcavity, with no
absorption by the material, the quality factor $Q$ is mainly
determined by the reflection loss at the interface between the
surface defect rods and the mirror rods .

\begin{figure}[h]
\includegraphics[width=5.0 in]{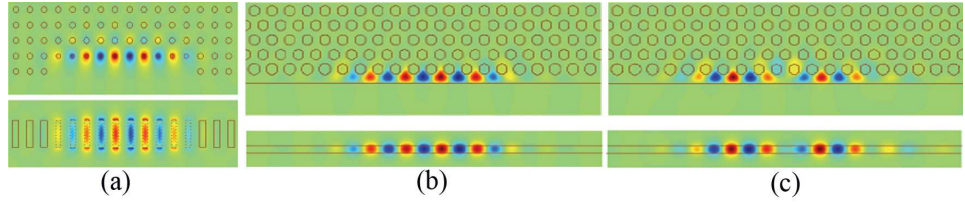}
\caption{\label{Fig3} (Color online) (a) Field cross sections for
the surface resonant mode in the square PhC slab cavity, showing
the $z$ component of the electric field.  (b) and (c) Field cross
sections for the two surface resonant modes [$\omega_b=0.2816
(a/\lambda)$ and $\omega_c=0.2841 (a/\lambda)$] in the triangular
PhC slab cavity, showing the $z$ component of the magnetic field.
Solid lines are outlines of the rods/holes/slabs.}
\end{figure}

Quality factors and resonant frequencies of surface resonant modes
are analyzed by the 3D FDTD method \cite{TafloveFDTD} with a
boundary treatment of perfectly matched layers
\cite{Berenger1994P1}. We excite the surface mode with a Gaussian
pulse and then monitor the radiative decay of the field. The
frequencies $\omega$ and quality factors $Q$ of the resonant modes
are calculated using a combination of FDTD techniques and Pad\'{e}
approximation with Baker's algorithm \cite{Guo2001P1}. In Fig.
\ref{Fig3} (a), we show the electric field cross sections of a
high-$Q$ surface resonant mode for the square PhC slab cavity with
the length of $L=11a$. Figure \ref{Fig3} (a) depicts horizontal
and vertical cross sections of the electric-field $z$ component at
the center plane of third axis ($z=0$) and at the plane center of
the surface defects, where the dotted line represents the position
of the surface defect rods with $R_d=0.15a$. The mode has an
angular frequency $0.3599 (a/\lambda)$ and quality factor $Q =
1.24\times 10^4$, in which $Q_\perp = 8.93 \times 10^4$ and $Q_\|
= 1.44 \times10^4$ ($1/Q=1/Q_\perp+1/Q_\|$). It can be clearly
seen from Fig. \ref{Fig3} (a) that the resonant mode is really
governed by wave vector $k=\pi/a$, which is consistent with what
we analyzed above.

\begin{figure}[h]
\includegraphics[width=8 cm]{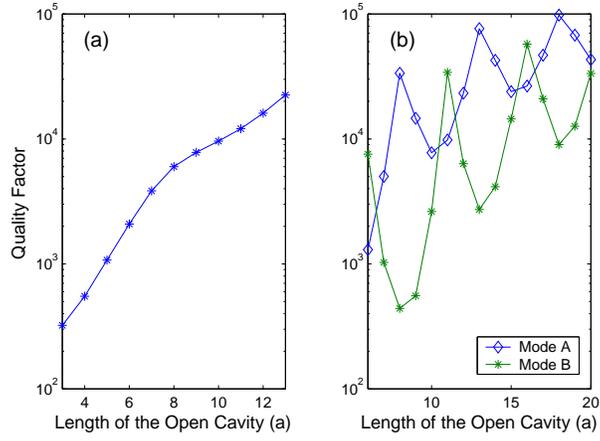}
\caption{\label{Fig4} (Color online) The quality factor (Q) is
plotted as a function of the length (L) for the microcavity in (a)
a square PhC slab and (b) a triangular PhC slab. Corresponding
structures are illustrated in Fig. 2. }
\end{figure}

From Fig. \ref{Fig3} (a), one can see that the electric field is
almost concentrated in the centers of the defect rods. We believe
that the reflection phase shift $\Delta\phi$ is close to $\pi$,
which is roughly independent of the cavity length $L$. Therefore,
the cavity modes can be related to surface modes which satisfy
$k=N\pi/L$, ($k\leq\pi/a$ and $N$ is an integer). However, our
results also show that the quality factor of the resonant mode
corresponding to $k=\pi/a$, i.e., zero-group-velocity surface
mode, is order of magnitude larger than those of cavity modes with
other $k$ vectors. $Q$ values for such a cavity mode with
different cavity length are shown in Fig. \ref{Fig4} (a). It can
be seen from Fig. \ref{Fig4} (a) that $Q$ gradually increases as
the cavity length becomes larger. Our results also show, with
increasing of the cavity length,  that the angular frequency of
the surface resonant mode converges to $\omega_0$, corresponding
to that with the zero-group-velocity surface mode in the PhC slab.


Let us next consider a triangular PhC slab with air holes
extending through a high-index ( $\varepsilon=11.56$)
finite-height dielectric slab, the top view of which is shown in
the inset of Fig. \ref{Fig1} (b). The holes have a radius of
$0.30a$, while the high-index slab is of thickness $0.6a$. The
distance between right boundary and the centers of the first right
holes is $d=\sqrt{3}a/2$. The dispersion relation for such a
triangular PhC slab transverse electric (TE) surface modes is
shown in Fig. \ref{Fig1} (b), where the shadow regions are the
projected band structure for TE modes. One can see from Fig.
\ref{Fig1} (b) that there exist three surface mode curves in the
triangular PhC slab. For simplicity, we only consider the angular
frequency between $0.26 (a / \lambda)$ and $0.30 (a / \lambda)$,
where there is one surface mode in a large region. Two
zero-group-velocity surface modes (Mode A with $k_1=\pi /a$ and
Mode B with $k_2=0.71 \pi /a$) can be found in our considered
frequency region. Now, imagine an optical cavity composed by such
a 2D triangular PhC slab, as shown in Fig. \ref{Fig2} (b). The
reflecting mirrors in the $y$ direction are introduced by
enlarging the radius of surface holes to $0.33a$, together with
the dielectric background. Using the method mentioned above, we do
find two high-quality-factor surface resonant modes with the
cavity length of $L = 11a$. One of the modes (corresponding to
Mode A) has a resonant angular frequency $0.2816 (a/\lambda )$ ,
$Q=9.81 \times 10^3$ and another (corresponding to Mode B) is
$\omega = 0.2841 (a/\lambda )$ , $Q=3.41 \times 10^4$. Figures
\ref{Fig3} (b) and \ref{Fig3} (c) show the magnetic-field cross
sections for the two resonant modes with the frequency of $0.2816
(a/\lambda )$ and $0.2841 (a/\lambda )$, respectively. The upper
and bottom figure represent the horizontal and vertical cross
sections of the magnetic field at the center plane of third axis
($z=0$) and at the vertical $yz$ plane (the distance between
surface boundary and the plane is $0.3a$), respectively.


Our results show that, when the cavity length increases, the
frequencies of the resonant modes converge to those two surface
modes with zero-group velocity. However, $Q$ values fluctuate as
the cavity length $L$ increases, as shown in Fig. \ref{Fig4}(b),
where the diamond line and the asterisk line represent the results
for Modes A and B, respectively. As described in Ref.
\cite{Ibanescu2005P1}, $Q$ should be enhanced when
$\Phi=2kL+2\Delta\phi$ is a multiple of $2\pi$. Due to the
complicated reflection interfaces, as seen in Fig. \ref{Fig2}(b),
the reflection phase shift $\Delta\phi$ is not close to $\pi$ for
the both modes, different from the case of the square lattice.
Furthermore, for the cavity length with numbers of lattice
constant, $2kL$ is not anymore close to a multiple of $2\pi$ for
the Mode B with $k=0.71\pi/a$. Peaks of $Q$ appear when $\Phi$ is
close to a multiple of $2\pi$. The positions of peaks for the Mode
A are different with those for the Mode B, mainly arisen from the
different wave vectors for the two resonant modes. $Q_A$ reaches
at $9.78 \times10^4$ when $L=18a$ and $Q_B$ is $5.73 \times10^4$
when $L=16a$. If the cavity length is further increased, $Q$
values will again fluctuate but in average increasing.

In this letter, we study optical microcavities based on surface
modes in PhCs. High quality factors can be obtained for
microcavities based on zero-group-velocity surface modes. We also
study the influence of the microcavity length to the quality
factor and resonant frequency. It has been shown that, with
increasing of the cavity length, the quality factor is gradually
enhanced and resonant angular frequency converges to $\omega_0$,
corresponding to that for the zero-group-velocity surface mode in
PhCs. The number of the resonant modes with high quality factors
is mainly determined by the number of surface modes with
zero-group velocity. It is well known that surface plasmons exist
on a metal-dielectric interface. We believe similar ideas can be
applied to a metallic structure to obtain surface resonant modes.

This work was supported by the Swedish Foundation for Strategic
Research (SSF) on INGVAR program, the SSF Strategic Research
Center in Photonics, and the Swedish Research Council (VR) under
Project No. 2003-5501.


\newpage
\section{Figure captions}
\textbf{FIG. 1}. (Color online) (a) The surface band structure for
the (10) surface of a square PhC slab of dielectric rods in air,
the radius of rods is $0.20a$ ($a$ is the lattice constant), the
height of the rods is $2.0a$, and the dielectric constant of the
rods is $11.56$. The radius of the rods at the surface is $0.15a$.
(b) The surface band structure for the $\Gamma K$ surface of a
triangular PhC slab of air holes extending through a high-index
($\varepsilon=11.56$) finite-height dielectric slab, the radius of
holes is $0.3a$, and the high-index slab is of thickness $0.6a$.
The distance between the centers of first right holes and the
right boundary of the dielectric is $d=\sqrt{3}a /2$.

\textbf{FIG. 2}. (Color online) (a) Top view of the microcavity
composed by a 2D square PhC slab. (b) Top view of the microcavity
composed by a 2D triangular PhC slab. The length of the cavities
is denoted by $L$.

\textbf{FIG. 3}. (Color online) (a) Field cross sections for the
surface resonant mode in the square PhC slab cavity, showing the
$z$ component of the electric field.  (b) and (c) Field cross
sections for the two surface resonant modes [$\omega_b=0.2816
(a/\lambda)$ and $\omega_c=0.2841 (a/\lambda)$] in the triangular
PhC slab cavity, showing the $z$ component of the magnetic field.
Solid lines are outlines of the rods/holes/slabs.

\textbf{FIG. 4}.  (Color online) The quality factor (Q) is plotted
as a function of the length (L) for the microcavity in (a) a
square PhC slab and (b) a triangular PhC slab. Corresponding
structures are illustrated in Fig. 2.

\newpage
\clearpage

\newpage
\clearpage

\newpage
\clearpage

\newpage
\clearpage

\end{document}